\documentclass{sig-alt-gov2}

\usepackage[draft]{hyperref}
\usepackage{listings}
\usepackage{courier}

\begin{document}

\conferenceinfo{JCDL'06,} {June 11--15, 2006, Chapel Hill, North Carolina, USA.}
\CopyrightYear{2006}
\crdata{1-59593-354-9/06/0006}

\title{An Architecture for the Aggregation and Analysis\\
	 of Scholarly Usage Data.}

\numberofauthors{2}

\author{
\alignauthor Johan Bollen\\
	\affaddr{Digital Library Research \& Prototyping Team}\\
       	\affaddr{Los Alamos National Laboratory}\\
       	\affaddr{Los Alamos}\\
       	\affaddr{NM, 87545}\\
       	\email{jbollen@lanl.gov}
\alignauthor Herbert Van de Sompel\\
	\affaddr{Digital Library Research \& Prototyping Team}\\
       	\affaddr{Los Alamos National Laboratory}\\
       	\affaddr{Los Alamos}\\
       	\affaddr{NM, 87545}\\
       	\email{herbertv@lanl.gov}
}

\date{January 2006}

\maketitle

\begin{abstract}
Although recording of usage data is common in scholarly information services, its exploitation for the creation of value-added services remains limited due to  concerns regarding, among others, user privacy, data validity, and the lack of accepted standards for the representation, sharing and aggregation of usage data. 
This paper presents a technical, standards-based architecture for sharing usage information, which we have designed and implemented.  In this architecture, OpenURL-compliant linking servers aggregate usage information of a specific user community as it navigates the distributed information environment that it has access to.  This usage information is made OAI-PMH harvestable so that 
usage information exposed by many linking servers can be aggregated to facilitate the creation of value-added services with a reach beyond that of a single community or a single information service. 
This paper also discusses issues that were encountered when implementing the proposed approach, and it presents preliminary results obtained from analyzing a usage data set containing about 3,500,000 requests aggregated by a federation of linking servers at the California State University system over a 20 month period.
\end{abstract}

\category{H.3.7}{Information storage and retrieval}{Digital Libraries}
\category{H.3.3}{Information Storage and Retrieval}{information Search and Retrieval}

\terms{Digital Libraries, usage data, architecture, standards, aggregation, analysis, OAI-PMH, OpenURL}


\section{Introduction}

Applications of usage data have altered the landscape of commercial information services. A user-driven revolution is underway in which end-user services are no longer solely based on top-down design decisions, but have come to prominently include the analysis of user actions and preferences. The introduction of a wide range of successful evaluation and recommendation services which rely in some shape or form on the analysis of usage data has fundamentally changed how users interact with online services and is having a profound effect on their behavior and preferences \cite{longta:anderson2005}.

It is natural that users of scholarly information services expect the same kind of functionality that is now common in the Web-based retail domain.  But the applicability of usage data in the scholarly realm goes beyond that of recommendation services, and also includes collection development support \cite{evalua:bollen2002}, derivation of metrics to assess the quality and impact of scholarly communication units \cite{altern:bollen2005}, and trend analysis \cite{bursty:kleinberg2002,detect:bollen2003}.  Especially for the latter two applications, the large-scale aggregation of usage data across information services and communities is essential, as the derivation of global measures of impact and the identification of global trends requires usage data that is representative at a global scale.



For that reason, a number of recent initiatives have focused on making usage data of scholarly information services available and on promoting its application in the domain of scholarly support services.  An XML-based format for the representation of Digital Library usage data was proposed by Goncalves (2002) \cite{xmllog:goncalves2002}, and a log archive was created to which usage logs could be posted, and later, downloaded. Unfortunately, neither the proposed format nor the log archive have been widely adopted. An OAI-PMH based solution for the representation and aggregation of usage data was proposed by Van de Sompel (2002) \cite {using:sompel2003}.  Although, in hindsight, the proposed format for the exposure of usage data seems to have been inspired by a specific application, that effort still provides the foundation for the work reported here. The COUNTER project\footnote{COUNTER: Counting Online Usage of NeTworked Electronic Resources, \scriptsize{\url{http://projectcounter.org/}}} has led to the specification of a format for recording journal-level usage statistics that information services can use to reliably share usage data with subscribing institutions. The {SUSHI} project\footnote{SUSHI: Standardized Usage Statistics Harvesting Initiative, \scriptsize{\url{http://www.library.cornell.edu/cts/elicensestudy/ermi2/sushi/}}} aims to automate the exchange and aggregation of COUNTER usage statistics by means of a dedicated set of web services and protocols. And, the IRS project\footnote{IRS: Interoperable Repository Statistics, \scriptsize{\url{http://irs.eprints.org/}}} focuses on usage information recorded by Institutional Repositories (IR) and aims to identify which parameters should be recorded uniformly across IRs, how such parameters can be derived from IRs and how the results can be shared for aggregation.

The remainder of this paper is organized as follows: Section 2 discusses the basic architecture of the proposed solution for the aggregation of usage data, and it details the manner in which the solution builds on accepted standards; to illustrate the potential of the proposed solution, Section 3 provides insights in prototype services that were developed on the basis of a large aggregated usage data collection; and Section 4 discussed issues that were encountered when implementing the proposed solution.
	
\section{Architecture}

This section outlines a technical, standards-based architecture for recording, representing, sharing and mining usage information of scholarly information services.  OpenURL-compliant linking servers play an important role in the proposed solution, as they naturally aggregate the navigations of a specific user community across the distributed information services that are available to them.  As will be discussed in detail in the remainder of this section, the following four phases can be distinguished in the proposed log harvesting architecture:


\begin{enumerate}
\item {\bf Intra-Institutional Aggregation of Usage Data} Usage events generated by users of a specific institution as they navigate their distributed scholarly information environment are recorded by the institutional linking server.
\item {\bf Exposure of Institutional Usage Data} The institutional usage data recorded by the linking server is exposed through an OAI-PMH-compliant log repository in which each event is represented as XML ContextObjects. 
\item {\bf Inter-Institutional Aggregation of Usage data} OAI-PMH harvesters collect the usage data from a variety of OAI-PMH-compliant institutional log repositories.
\item {\bf Service Provision} Value-added services are implemented based on the aggregated usage data collection.
\end{enumerate}

\subsection{Intra-Institutional Aggregation of Usage Data}

Starting around 2001, scholarly information services have in great numbers begun to support the concept of context-sensitive services \cite{openli:vandesompel2001} by implementing the OpenURL 0.1 specification\footnote{\scriptsize{\url{http://www.exlibrisgroup.com/sfx\_openurl\_syntax.htm}}},
while academic and research libraries have increasingly installed the linking servers that are required to provide localized services to their user base.

\begin{figure}[ht!]
\begin{center}
\includegraphics[width=2.5in]{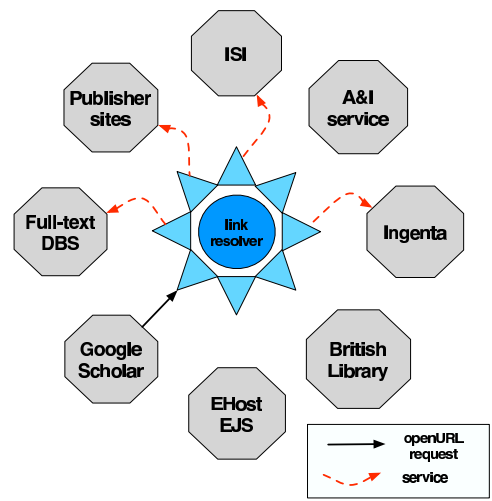}
\end{center}
\caption{\label{localaggregation} Linking servers are well-positioned to capture usage data.}
\end{figure}

Fig. \ref{localaggregation} shows the information environment of a particular user community as it could exist for an academic institution. It shows the many distributed scholarly information services that are accessible to that user community, and it shows an institutional linking server as a central hub in this information environment.  In the context-sensitive service environment enabled by OpenURL and linking servers, an information service, such as Google Scholar, inserts an OpenURL for every reference to a scholarly work that it presents to a user, for example, as a search result.  This OpenURL is an HTTP GET request carrying metadata that are essential to identify the referenced work. It points to the linking server of the users' institution which contains a rule engine powered by a knowledge database that is typically maintained by the user's institutional library.  Given an incoming OpenURL request, and through consultation of its localized rules and knowledge database, the linking server returns a list of services pertaining to the referenced work to the user.  Those services typically point into other information services available in the users' distributed information environment, such as e.g. Ingenta, ISI, Publishers sites and Full-text DBS  as shown in Fig. \ref{localaggregation}.

The central position held by a linking server in the distributed scholarly information environment of a specific user community makes it particularly appealing as the source of usage information.  Indeed, a linking server logs the OpenURL requests of all users of the community originating from many of the available distributed information sources. As such it de-facto aggregates usage information at the level of the community, and internally represents the usage information in a normalized manner.

A particular and unique advantage of linking servers usage data is the ability to track sequences of requests across a variety of information services. Indeed, in Fig. \ref{localaggregation}, the linking server is aware of the fact that the user was navigating Google Scholar, and requested services from the linking server pertaining to a referenced work.  The linking server also knows which of those services was chosen by the user.  And, if that chosen service led into another OpenURL-compliant information service, e.g .Ingenta in Fig. \ref{localaggregation}, and if the user again requested services from the linking server pertaining to a work referenced there, the linking server would again be aware. Such a sequence of requests can be recorded by the linking server and hence exploited by click-stream based methods of log analysis \cite{pageco:kothari2003,effect:mobasher2001} to reveal temporal trends in user behavior and recommending items which are often accessed in a particular sequence.  Such temporal patterns would be very difficult, if not impossible, to reconstruct from the aggregation of the logs obtained from each of the individual information services in the users' environment.

\subsection{The OpenURL ContextObject for the interoperable representation of usage data}

Each brand of available OpenURL-compliant linking server probably stores its usage information in a proprietary manner.  However, when the goal is to share usage information across a federation of heterogeneous linking servers, support for a common representation format becomes important.

\begin{figure}[ht!]
\begin{center}
\includegraphics[width=3.2in]{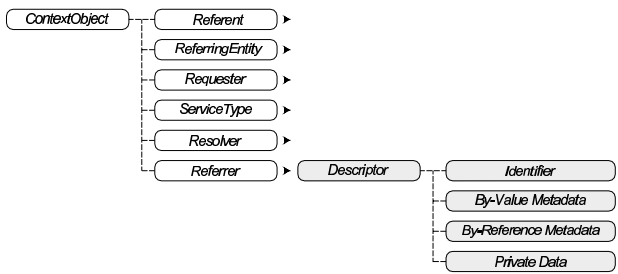}
\end{center}
\caption{\label{contextobject} Structure of OpenURL ContextObject.}
\end{figure}

The NISO/ANSI Z39.88-2004 Standard "The OpenURL Framework for Context-Sensitive Services"\footnote{\scriptsize{\url{http://library.caltech.edu/openurl/default.htm}}} is a powerful generalization of the context-sensitive service concepts that were at the basis of the definition of OpenURL 0.1. 
At the core of the standard is the notion of the ContextObject.  The ContextObject is an abstract data structure that encapsulates six entities that are involved in the fulfillment of a context-sensitive service request. The structure of the OpenURL ContextObject is shown in Fig. \ref{contextobject}. At the core of the ContextObject is the {\it Referent}; it is the actual subject of the service request that the ContextObject encodes, i.e. the service request pertains to the {\it Referent}. The {\it Requester} is the agent that requests the service pertaining to the {\it Referent}. The {\it ServiceType} specifies the type of service that is requested.  The ContextObject furthermore contains the {\it ReferringEntity}, i.e, the entity that references the {\it Referent}, the {\it Resolver} which is the target of a service request, and the {\it Referrer} which is the system that generated the ContextObject.

 Each entity of the ContextObject can be described by means of a combination of identifiers, metadata and private data.  To allow for a controlled deployment of applications based on the OpenURL Standard, the OpenURL Registry\footnote{\scriptsize{\url{http://www.openurl.info/registry}}} provides the capability to register identifier namespaces and Metadata Formats that are used in OpenURL Applications.  The abstract ContextObject data structure can be instantiated using different serializations, and both a Key/Encoded-Value pair serialization and an XML serialization have been defined as part of the NISO standardization effort. A service request pertaining to a {\it Referent} is achieved by transporting a serialized ContextObject with the {\it Referent} at its core towards a {\it Resolver}.

In the perspective of the NISO/ANSI Z39.88-2004 Standard, linking servers as described above have become a special type of {\it Resolver}, namely {\it Resolver} that supports a specific OpenURL Application known as the San Antonio Profile\footnote{\scriptsize{\url{http://www.openurl.info/registry/docs/pdf/SanAntonioProfileLevel1-2004.pdf}}}$^,$\footnote{\scriptsize{\url{http://www.openurl.info/registry/docs/pdf/SanAntonioProfileLevel2-2004.pdf}}}.   Also, according to the new standard, a service request targeted at a linking server is the transportation of a ContextObject with a description of the referenced work at its core (the {\it Referent}) towards the linking server (the {\it Resolver}).  Hence, since a ContextObject is the embodiment of a service request aimed at linking servers, the {ContextObject} also provides an appropriate data structure for the representation and sharing of usage information recorded by linking servers.


In order to illustrate the mapping between a service request issued to a linking server and the ContextObject data structure, it is worthwhile pointing out that each individual usage event can, in essence, be described by a triplet consisting of:
\begin{description}
	\item[What] The item for which the usage was recorded, e.g. a journal article. 
	\item[Who]  The originator of the event, e.g. the user
	\item[When] The time at which the event occurred, i.e. the event's timestamp
\end{description}

In the proposed usage log representation technique, a usage event is defined as an individual OpenURL-compliant service request targeted at a linking server.  Such a usage event is represented as an individual ContextObject according to the mapping described in the remainder of this paragraph. The {\bf what} and {\bf who} components of the triplet can readily be mapped to the {\it Referent} and {\it Requester} entities of the ContextObject, respectively.  Moreover, as can be seen in Table \ref{COmapping}, the ContextObject allows for the inclusion of descriptions of other entities that are relevant to a service request and hence to the downstream exploitation of the represented usage data: the {\it Resolver}, the {\it ServiceType}, the {\it Referrer}, and the {\it ReferringEntity}. These mappings are independent of the serialization format of the ContextObject. In order to include the {\bf when} component of the triplet, a choice for the XML serialization of the ContextObject must be made as that provides an administrative information element, namely the \texttt{timestamp} attribute, which indicates when the ContextObject was generated, which in the context of this application is the moment at which the service request was issued (see Table \ref{openurlCO}).  Moreover, when aiming at the global sharing of usage information it is important to be able to unambiguously identify each event recorded by a linking server.  To that end, each event represented by a ContextObject is accorded a globally unique UUID [9] by the linking server, which can be conveyed using another administrative element of the XML ContextObject, namely the  \texttt{identifier} (see Table \ref{openurlCO}).

The ability to express all the entities that pertain to a service request in a standard-based, self-contained data structure is quite appealing in light of the need to share usage data across the boundaries of information services and communities.  Because the proposed mappings are done with the aim of sharing usage information across a federation of linking servers, the choice for the XML ContextObject format seems logical instead of restrictive. Moreover, the XML ContextObject format allows entities of the ContextObject to be described using more than one Metadata Format, as such allowing for very expressive descriptions whenever possible or appropriate.

\begin{table}[ht!]
\begin{tabular}{lp{5.5cm}}
{\bf Entity}		&		{\bf Mapped usage data}\\\hline
{\it Referent}				&		Item used, e.g. journal article\\
{\it ReferringEntity}			&		Entity that referenced the {\it Referent}\\
{\it Requester}				&		User identification\\
{\it ServiceType}			&		Type of usage, e.g. "full-text request"\\
{\it Resolver}				&		Linking server\\
{\it Referrer}				&		Information system that generated the ContextObject, i.e. the system that the user was navigating when issuing a service request to the linking server\\\hline
\end{tabular}
\caption{\label{COmapping} Mapping of usage data  to the ContextObject data structure}
\end{table}

Table \ref{openurlCO} shows an example of linking server usage data encoded according the described mapping; for brevity XML Namespace declarations have been omitted:
\begin{itemize}
\item The root  \texttt{context-object} element has two attributes, the  \texttt{timestamp} and the  \texttt{identifier}, with semantics as described above.
\item The {\it Referent} is a journal article that is being described by both an identifier and by metadata.  The identifier is a DOI identifier expressed as a URI following the "info" URI scheme\footnote{\url{http://info-uri.info/}}, while the metadata is compliant with the XML Metadata Format to describe journal articles.  This Metadata Format is registered in the OpenURL Registry\footnote{\scriptsize{\url{info:ofi/fmt:xml:xsd:journal}}} along with its XML Schema definition\footnote{\scriptsize{\url{http://www.openurl.info/registry/docs/xsd/info:ofi/fmt:xml:xsd:journal}}}. The OpenURL Registry contains XML Metadata Format definitions for other types of scholarly works including dissertations, books, and patents.  Metadata Format definitions for other types of works, for example datasets, can be registered, and, for flexibility, the OpenURL Standard allows for the use of unregistered Metadata Formats.
\item The {\it Requester} is described by means of an identifier which in this example consists of the IP address of the user's computer represented using an ad-hoc URN scheme.  Typically, the IP address is the only information that is available for a {\it Requester}.  However, if more information, such as an EduPerson user record\footnote{\scriptsize{\url{http://www.educause.edu/eduperson/}}}, would be available, it could be expressed using XML and hence could be included as a metadata description of the {\it Requester}.  Similarly, the inclusion of session information as a more expressive proxy to the {\it Requester} than the IP address is possible through the definition of a Metadata Format or identifier scheme. However, due to privacy concerns it is likely that less rather than more  {\it Requester} information would be made available when sharing usage data beyond the boundaries of an institution, or that such information would become encrypted or anonymized.
\item The {\it ServiceType} is described by means of a Metadata Format registered in the OpenURL Registry\footnote{\scriptsize{\url{info:ofi/fmt:xml:xsd:sch\_svc}}}.  This Metadata Format allows for the expression of one or more services that were actually requested by the user from the linking server about the scholarly item that is the {\it Referent}.  The example indicates that the user requested the full-text of the article; other services that are expressible using this Metadata Format include "abstract", "citation" and "holding".  
\item The {\it Resolver} is described by means of an identifier, which is the HTTP address of the linking server at which the service request was targeted. Again, it can easily be imagined that this information would be encrypted or anonymized if usage information is shared beyond an institution.
\item The {\it Referrer} is described by means of an identifier using the registered "sid" namespace of the info URI scheme.  The identifier indicates that the user was navigating the Scopus service of Elsevier Science when requesting services from the linking server.
\end{itemize}

\lstset{morecomment=[l],emph={Resolver,Referrer,object,type,ctx:metadata,ctx:context,ctx:referent,ctx:requester,ctx:referent,ctx:service}, emphstyle=\bf}

\begin{table*}[ht!]
\begin{scriptsize}
\begin{lstlisting}[language=XML,frame=ltrb,mathescape=true,commentstyle=\bf]
<?xml version="1.0" encoding="UTF-8"?>
<ctx:context-object
  timestamp="2005-11-11T17:45:08Z"                                         <!-- event date and time -->
  identifier="urn:UUID:58f202ac-22cf-11d1-b12d-002035b29062">            <!--  global event ID -->
<ctx:referent>                                                           <!-- referent data -->
    <ctx:identifier>info:doi/10.1016/j.ipm.2005.03.024</ctx:identifier>                  <!-- referent identifier -->
    <ctx:metadata-by-val>				                   <!-- referent metadata -->
      <ctx:format>info:ofi/fmt:xml:xsd:journal</ctx:format>
      <ctx:metadata>
        <jou:journal>
          <jou:atitle>Toward alternative metrics of journal impact</jou:atitle>
        <jou:jtitle>Information Processing and Management</jou:jtitle>   $\cdots$   
 </ctx:metadata> 
  </ctx:referent>
$\cdots$
  <ctx:requester>				      		           <!-- requester ID -->
      <ctx:identifier>urn:ip:63.236.2.100</ctx:identifier>
  </ctx:requester>
$\cdots$
  <ctx:service-type>				      		           <!-- type of request -->
   <ctx:metadata-by-val>				                   <!-- referent metadata -->
      <ctx:format>info:ofi/fmt:xml:xsd:sch_svc</ctx:format>
        <ctx:metadata>
         <full-text>yes</full-text> $\cdots$
        </ctx:metadata> 
  </ctx:service-type>
  <ctx:resolver>				      		          <!-- resolver ID -->
      <ctx:identifier>http://sfx.example.org/menu</ctx:identifier>
  </ctx:resolver>  
  <ctx:referrer>				      		          <!-- referrer ID -->
      <ctx:identifier>info:sid/elsevier.com:scopus</ctx:identifier>
  </ctx:referrer>  
</ctx:context-object>
\end{lstlisting}
\caption{\label{openurlCO}Abbreviated sample demonstrating the representation of usage data as OpenURL ContextObjects}
\end{scriptsize}
\end{table*}

\subsection{Inter-Institutional Aggregation of Usage Data}

Usage data pertaining to the scholarly information collection of a specific institution is a valuable asset for those institutions that choose to record and exploit it.  For example, it allows management to track usage as it occurs and to make accurate and community-driven collection development decisions \cite{evalua:bollen2002}. It can also be used in recommender services to enhance the discovery capabilities of users.

However, increased value of usage data can be realized when it is aggregated over a large number of information sources and communities such that a representative sample of the activities of the scholarly community (or a well-defined subset thereof) is obtained. If such aggregation can be achieved, applications can be imagined that discover, analyze and predict trends in the scholarly endeavor; recommender systems can be built that are based on the activities of a global scholarly user base; and a new generation of usage-based impact and quality metrics can be defined, deployed and used to balance the monopoly of the citation-based ISI Impact Factor \cite{readin:darmoni2002,altern:bollen2005}.

To allow for the emergence of large-scale collections of usage data, mechanisms for exchange and aggregation need to be devised. In the proposed approach, linking servers are used as {\it intra-institutional} aggregators of usage information.  In order to allow for the {\it inter-institutional} aggregation of usage information, an OAI-PMH based technique is proposed in which an OAI-PMH repository with the following core properties exposes the usage logs of an institutional linking server:

\begin{itemize}
\item Contained records are XML ContextObjects only. Each ContextObject represents an event recorded by the linking server as explained above.
\item The {\it identifier} used by the OAI-PMH is the globally unique UUID that unambiguously identifies a linking server event; it is the same as the value of the \texttt{identifier} attribute to the root element of the XML ContextObject.
\item The {\it datestamp} used by the OAI-PMH is the datetime the event was uploaded to the log repository.  Because it is expected that the log repository will be a derivative of the logs as stored by the linking server, the OAI-PMH {\it datestamp} does not coincide with the datetime of the event as provided in the \texttt{timestamp} attribute to the root element of the XML ContextObject. It should be noted that the {\it datestamp} of a record never changes, as an event will never be updated once it has been recorded and uploaded to the log repository.  As a result, once harvested by a usage data aggregator, a record will not have to be re-harvested.
\item The only supported {\it metadata format} is the XML ContextObject Format (with metadataPrefix resolver\_logs), registered in the OpenURL Registry\footnote{\scriptsize{\url{info:ofi/fmt:xml:xsd:ctx}}} along with an XML Schema definition\footnote{\scriptsize{\url{http://www.openurl.info/registry/docs/xsd/info:ofi/fmt:xml:xsd:ctx}}}.
\item The harvesting {\it granularity} can be either at the day-level or the seconds-level.
\item No {\it set} structure is provided. 
\end{itemize}

Table \ref{oai_record} shows an example of an OAI-PMH record that contains the ContextObject of Table \ref{openurlCO}.\\

This OAI-PMH-based approach allows harvesters to recurrently collect usage data as recorded by institutional linking severs, and to compile a usage data collection with a global or regional reach. The creation of aggregated collections with a different focus, such as a discipline-specific aggregation, would either require post-processing of harvested logs, or the introduction of conventions regarding discipline-oriented {\it set} structures at the level of the log repositories.  The latter has proven to be problematic for OAI-PMH repositories that expose Dublin Core metadata, and, given their noisy nature, may turn out to be unrealistic to achieve for usage data.  

\begin{figure}[ht!]
\begin{center}
\includegraphics[width=3.2in]{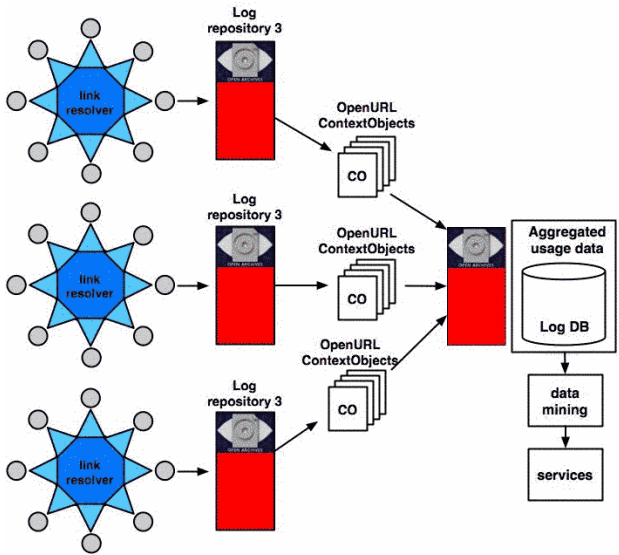}
\end{center}
\caption{\label{bxarch} Exposing and harvesting linking server logs using OAI-PMH and OpenURL ContextObjects.}
\end{figure}

\begin{table}[h!]
\begin{scriptsize}
\begin{center}
\begin{lstlisting}[emph={record, header,identifier,timestamp,datestamp, metadata,ctx,context,object},language=C,frame=ltrb,mathescape=true]
<oai:record>
<oai:header>
  <oai:identifier>
urn:UUID:58f202ac-22cf-11d1-b12d-002035b29062
  </identifier>
  <oai:datestamp>
    2005-11-12T21:21:51Z
  </oai:datestamp>
</oai:header> 
<oai:metadata>
  <ctx:context-object
  identifier="urn:UUID:58f202ac-22cf-11d1-b12d..."
  timestamp="2005-11-11T17:45:08Z">
   <ctx:referent>                                                                        
     <ctx:identifier>
     info:doi/10.1016/j.ipm.2005.03.024
     </ctx:identifier>
     $\cdots$
   </ctx:referent>                                                                        
    $\cdots$
  </ctx:context-object>
</oai:metadata>
</oai:record>
\end{lstlisting}
\caption{\label{oai_record} Sample OAI-PMH record containing OpenURL ContextObject}
\end{center}
\end{scriptsize}
\end{table}

\subsection{Service Provision}

Once usage data is aggregated across the boundaries of a single information service and a single institution, as facilitated by the aforementioned approach, services can be created on the basis of the aggregated usage data collection. A major attraction of the proposed approach is that many aggregators can emerge, each of which could experiment with the creation of yet unknown services by mining the usage data collection in a variety of ways.  As described in the following Section, our experiments have so far mainly focused on the creation of a recommender system and on the extraction of metrics that may eventually be attractive for the assessment of the quality and impact of scholarly works.

\section{Results}

As a proof of concept, the described architecture was implemented in conjunction with the most widely deployed commercial linking server, the SFX server from Ex Libris\footnote{\scriptsize{\url{http://www.exlibrisgroup.com/sfx.htm}}}.  To that end, an autonomous usage data add-on to the SFX linking server software with the following capabilities was implemented: 
\begin{enumerate}
	\item The add-on can ingest the linking server's usage data into a special purpose log database.
	\item The linking server's log database is exposed as an OAI-PMH-compliant log repository with characteristics as described in the above.
	\item The add-on contains an OAI-PMH usage data harvester which has the ability to collect usage data from remote OAI-PMH-compliant log repositories, and to merge the harvested usage data with the linking server's own usage data.  Care is taken not to re-expose usage data that was obtained through harvesting from remote linking servers.
\end{enumerate}

To gain experience with building services on the basis of a large log collection, usage data data was aggregated across the California State University (CalState) campuses. The CalState system is one of the largest systems of public universities in the US. It comprises 23 campuses and seven off-campus centers which combined have a population of 409,000 students and 44,000 faculty and staff. The CalState system has deployed SFX live since Fall 2002 and uses an SFX consortium model consisting of 23 SFX linking servers (one per campus) and 1 for shared resources (operated by the Chancellor's Office). For reasons of scale and its long-standing use of linking servers the CalState data offered a unique testing opportunity.  Our initial experiences with building services on the basis of a large usage data collection are discussed in the remainder of this section.

\subsection{Aggregated Usage Data Collection}

Usage data from 9 CalState linking servers were included for an initial analysis.  These 9 linking servers were selected because their IP-address distributions suffered the least from IP-address distortions caused mainly by the reliance on proxy servers when requesting services from a linking server.  When accessing a linking server through a proxy, the real IP-addresses of the {\it Requester} are obscured\footnote{Later modifications were made to replace the IP address with an anonymous session ID to avoid this issues}.

The 9 selected instances were Chancellor, CPSLO, Los Angeles, Northridge, Sacramento,  San Jose, San Marcos, SDSU, and finally SFSU. They represented the majority of linking server usage data in the CalState system.  Usage data had been recorded at these institutions in the period November 11th, 2003 (10:44 AM) to August 18th, 2005 (11:43PM).

This data set was aggregated and loaded into the aforementioned add-on's special purpose log databases.  The resulting collection consisted of 3,507,484 unique usage events.  A de-duplication process run on the basis of the identifiers and the metadata describing the {\it Referents} (documents) involved in the events, resulted in a total of 2,133,556 unique {\it Referents} in the data set; the set contained 167,204 unique {\it Requesters} when using the user agent's IP-address as the {\it Requester's} identifier. A majority, i.e.~67\%, of the {\it Referents} were journal articles.

\subsection{Mining item relationships from CalState usage data}

Usage data naturally consists of a temporally ordered but otherwise unconnected sequence of usage events. In order to perform more sophisticated, network-based methods of {\it Referent} ranking and to create recommender services able to link one {\it Referent} to the other, a network of item relationships needed to be extracted from the CalState usage data. It was therefore subjected to a technique similar to that employed by amazon.com which relates products if they have been purchased by the same users. As a corrolary we related pairs of {\it Referents} in the CalState usage data according to the degree to which they were consistently accessed by the same users. More details on this technique, which is highly related to collaborative filtering \cite{itemba:sarwar2001} and association rule learning \cite{effect:mobasher2001}, can be found in \cite{altern:bollen2005}.

This procedure resulted in  a network of {\it Referent} relationships represented by an {\it Referent} relationship matrix. Since usage is not bound by {\it Referent} type (journals, articles, etc) {\it Referents} were not differentiated in the network and therefore journals could e.g.~be related to articles and vice versa.  As a starting point for our preliminary data analysis we needed a more focused network. Therefore, a network of relationships between journals was generated by aggregating all journal article relationships between articles published in the same journal. The resulting journal relationship network pertained to a set of 45,554 journals for which 1,927,506 journal relationships were derived. Table \ref{matrix} shows a sample of the matrix representing this journal level network; each entry in the matrix reflects the strength of the relationship as inferred from the usage data collection.

\begin{table}
\begin{scriptsize}
\begin{center}
\begin{tabular}{l||c||p{0.10cm}p{0.1cm}p{0.1cm}p{0.1cm}p{0.1cm}p{0.1cm}p{0.1cm}p{0.1cm}p{0.1cm}p{0.1cm}}
					&	 	& 1		& 2		& 3		& 4		& 5		& 6		& 7		& 8		& 9		&	10		\\\hline\hline
DISS ABSTR A		&	1	& 0	& 0	& {\bf 51}	& 0  	& 0  	& 0	& 0	& 0	&{\bf 4}	&	0	\\
NY TIMES				&	2	& 0 	& 0	& {\bf 6}	& {\bf 48}  	& {\bf 5}  	& 0	& {\bf 11}	& 0	& 0	&	0	\\
DISS ABSTR B		&	3	& {\bf 47}	& {\bf 5}	& 0	& 0  	& 0  	& 0	& 0	& {\bf 5}	& {\bf 13}	&	{\bf 5}	\\
WALL STREET J		&	4	& 0	& {\bf 47}	& 0	& 0  	& 0  	& 0	& 0	& 0	& 0	&	0	\\
SCIENCE				&	5	& 0	& {\bf 6}	& 0	& 0  	& 0  	& {\bf 14}	& 0	& 0	& 0	&	0	\\
NATURE				&	6	& 0	& 0	& 0	& 0  	& {\bf 14}  	& 0	& 0	& 0	& 0	&	0	\\
NY TIMES MAG 		&	7	& 0	& {\bf 12}	& 0	& 0  	& 0  	& 0	& 0	& 0	& 0	&	0	\\
DEV PSYCH	 		&	8	& 0	& 0	& {\bf 5}	& 0  	& 0  	& 0	& 0	& 0	& 0	&	{\bf 11}	\\
PSYCH REP			&	9	& {\bf 4}	& 0	& {\bf 11}	& 0  	& 0  	& 0	& 0	& 0	& 0	&	0	\\
CHILD DEV			&	10	& 0	& 0	& {\bf 54}	& 0	& 0	& 0	& 0	&{\bf  10}	& 0	&	0	\\\hline
\end{tabular}
\caption{\label{matrix} Sample of journal usage matrix.}
\end{center}
\end{scriptsize}
\end{table}

\subsection{Usage impact ranking}

The Google search engine uses the PageRank algorithm to determine the impact of web pages on the basis of how often they are linked to by high impact web pages \cite{anatom:brin1998,inside:bianchini2005}. A page receiving many in-links from high-impact pages is assumed to be of high impact itself. Since a network of journal relationships has been established from the mentioned CalState usage data, its journals can now be ranked according to the same algorithm which has proven effective in web searches. The PageRank values calculated on the basis of usage-defined journal networks is referred to as Usage PageRank.

Table \ref{indegIF} lists the 10 highest scoring journals according to their Usage PageRank and their corresponding 2003 citation Impact Factors. The latter reflects the impact of a journal according to the frequency by which its articles are cited over a 2 year period, i.e. 2001 and 2002. A juxtaposition between the Usage PageRank and citation Impact Factors can therefore reveal how the impact of a journal in the CalState system deviates from the general scholarly community. Journals whose usage PageRank deviates strongly from the citation impact factor are therefore marked with a "$\star$".

These results indicate that usage and citation indicators of journal impact agree only for a number of top journals such as Nature and Science. However, for a large group of journals there exist significant deviations which correspond to what one could assume is the institutional research focus in the CalState system. In other words, the journals "Nature" and "Science" are equally important in the CalState community as they are in the general scholarly community, but the Journal of Advances in Nursing (J ADV NURS) is much more important in the CalState community than its general citation rates indicate.

The rankings on the basis of Usage PageRank offer the enticing possibility of more accurately pinpointing the dynamic preferences and interests of a particular user community, in this case CalState. In addition, they confirm earlier results obtained for the Los Alamos National Laboratory user community \cite{altern:bollen2005}. One could speculate that with increasing sample sizes these rankings could provide a global indication of the status of scholarly communication items for the entire scholarly community thereby augmenting existing citation-based methods.

\begin{center}
\begin{table}[ht!]
\begin{tabular}{ccp{1cm}r}
Rank	&	PRw	&	IF03		&	Journal Title						\\\hline
1		&	115.388		&	21.455	&	JAMA					\\
2		&	102.377		&	29.781	&	SCIENCE					\\
3		&	86.040		&	30.979	&	NATURE					\\
4		&	63.282		&	3.779	&	J AM ACAD CHILD PSY$\star$\\
5		&	61.474		&	7.157	&	AM J PSYCHIAT $\star$		\\
6		&	56.156		&	3.363	&	AM J PUBLIC HEALTH $\star$	\\
7		&	55.488		&	34.833	&	NEW ENGL J MED			\\
8		&	53.712		&	2.591	&	MED SCI SPORT EXER $\star$\\
9		&	40.17		&	0.998	&	J ADV NURS $\star$			\\
10		&	39.123		&	5.692	&	AM J CLIN NUTR $\star$		\\
\end{tabular}
\caption{\label{indegIF} Journal PageRank (PRw) in CalState usage data and 2003 citation Impact Factors (IF03).}
\end{table}
\end{center}

\subsection{Journal level interest mapping}
\begin{figure}[ht!]
\begin{center}
\includegraphics[width=3.2in]{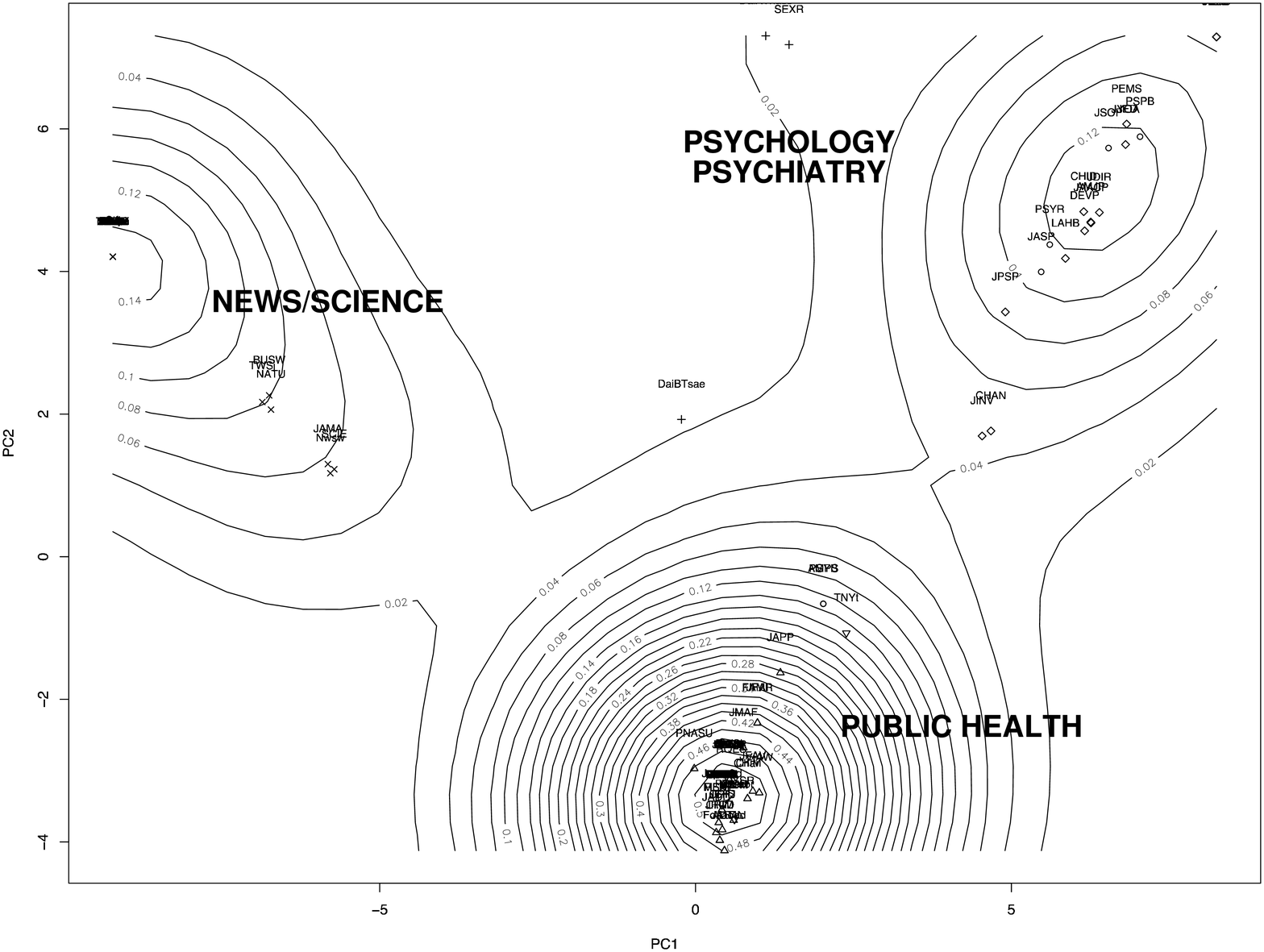}
\end{center}
\caption{\label{calstatePCA} Mapping of journals accessed in CalState system on the basis of usage.}
\end{figure}

The ranking of journals according to Usage PageRank offers an informative view of the CalState community's characteristics. A comparison with the 2003 citation Impact Factors highlights journals of particular interest. To describe the properties of this community in finer detail, a geographical mapping of journal usage relationships can be generated by a Principal Component Analysis \cite{princi:jolliffe2002}. Such a mapping places journals in a 2-dimensional location according to how similar or dissimilar their usage is. Fig. \ref{calstatePCA} shows the resulting mapping of journal similarities derived from the CalState usage data. The graph reveals three main clusters of interests namely a "news" (top left), "psychology" (top right) and "public health and policy" (mid bottom) cluster. This mapping forms a model of how CalState users interact with their information services and can thus serve as the basis for an analysis of user habits and interests. The fact that a meaningful structure emerges in this mapping indicates the validity and quality of the aggregated linking server usage data.

\subsection{Recommender Services}

The generated relationship networks encode which journals (or any type of {\it Referents}) are related in their usage to which others. They can therefore be used to recommend documents whenever a user expresses interest in a specific document (or set thereof). On the basis of the general {\it Referent} relationship matrices, a prototype recommender service was constructed which accepts a description of a journal or article (identifier and/or metadata) as input and then scans the relationship matrices for viable suggestions. 

\begin{figure}[ht!]
\begin{center}
\includegraphics[width=3.2in]{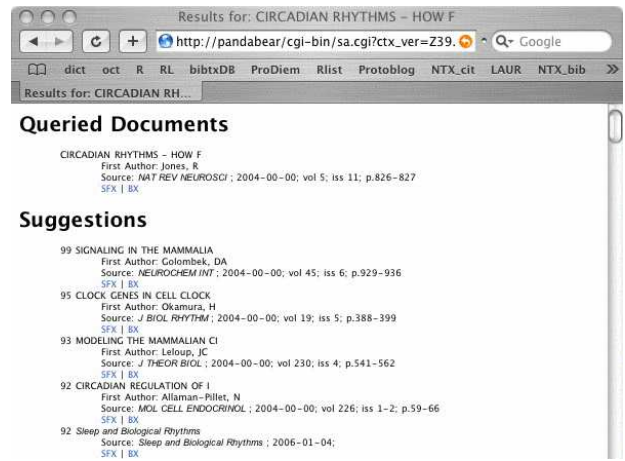}
\end{center}
\caption{\label{recomm} Generating usage-based recommendations}
\end{figure}

Fig. \ref{recomm} shows a screenshot of the implemented prototype. Table \ref{recomm1} and \ref{recomm2} show the results obtained for two recommendation requests. In the case of Table \ref{recomm1}, recommendations were requested for an article on the subject of "Circadian rhythms". Among the top ten results, we indeed find mostly articles strongly related to varying aspects of circadian rhythmes and the physiological aspects of biological clocks. Table \ref{recomm2} lists the results generated for a query relating to the issue of learning reading skills at an early age. Indeed, all top 10 ranked recommendations relate to education and schooling issues. Note that results are obtained on the basis of usage data, not on the basis of term extraction. For example, in Table \ref{recomm1} an article entitled "You talking to me?" is issued as a valid recommendation, even though none of its metadata items matches those of the query document.

Although these results do not represent a valid, quantitative analysis of the effectiveness of usage-based recommendations, they do serve as a promising pointer to the potential value of scholarly usage data for advanced end-user services. In fact, the principle of deriving recommender systems from usage data has already been widely validated in the literature \cite{itemba:sarwar2001,protot:hwang2003} and we expect scholarly usage data to be no exception.

\begin{table*}
\begin{center}
\begin{tabular}{cl}
\multicolumn{2}{c}{{\bf "R. Jones (2004) Circadian rhythms: How time flies. NAT REV NEUROSCI. 5(11), 826-827"}}\\
rank	&	recommendation\\\hline
1	&	DA Golombek (2004). Signaling in the mammalia.  NEUROCHEM INT 45(6), 929-936\\
2	&	H Okamura (2004). Clock genes in cell clock: Roles, Actions, and Mysteries. J BIOL RHYTHM 19 (5), 388-399\\
3	&	JC Leloup (2004).   Modeling the mammalian circadian clock: Sensitivity analysis... J THEOR BIOL 230(4), 541-562\\
4	&	N Allaman-Pillet. (2004) Circadian regulation of islet genes involved in insulin... MOL CELL ENDOC 226(1-2), 59-66\\
5	&	S Panda (2004). It'as all in the timing: Many Clocks, Many Outputs. J BIOL RHYTHM 19(5), 374-387\\
6	&	M Zatz (2004). You talking to me? J BIOL RHYTHM 19(4), 263-263\\
7	&	Jadwiga Giebultowicz (2004). Chronobiology: Biological Timekeeping. INT COMP BIOL 44(3), 266\\
8	&	M. Shermer (2004). None so blind. SCI AM  290(3), 42-42\\
9	&	H Kobayashi (2004). Effect of feeding on peripheral circadian rhythms... GENES CELLS 9(9), 857-864\\
10	&	R Sitruk-Ware (2004). New progestogens - A review of their effects. DRUG AGING 21(13). 865-883\\\hline
\end{tabular}
\end{center}
\caption{\label{recomm1} Usage-based recommendations for "Circadian rhythms" query.}
\end{table*}

\begin{table*}
\begin{center}
\begin{tabular}{cl}
\multicolumn{2}{c}{{\bf "R. Gersten (2003) Teaching reading to early language learners. EDUC LEADERSHIP 60 (7), 44-49"}}\\
rank	&	recommendation\\\hline
1	&	A Thompson (2002). May be we can just be friends. EDUCATIONAL THEORY 52 (3), 327-38\\
2	&	T Quiroga(2002). Phonological awareness in Spanish. J SCHOOL PSYCHOL 40(1), 85-111\\
3	&	S Linan-Thompson (2003). Effectiveness of supplemental reading instruction... ELEM SCHOOL J 103(3), 221-238\\
4	&	L Araujo (2002) The literacy development of kindergarten English-language... J RES CHILD EDUC 16(2), 232\\
5	&	L. Morris (2001). Going through a bad spell: what young ESL learnersÕ... CAN MOD LANG REV 58(2), 273-286\\
6	&	SO Dahlgren (1996). Theory of mind in non-retarded children... J CHILD PSYCHOL PSYCH 37(6), 759-763\\
7	&	EG Cohen (2002). Can groups learn? TEACH COLL REC 104(6), 1045-1068\\
8	&	D Freeman (2000). Meeting the needs of English language learners. TALKING POINTS 12 (1), 2-7\\
9	&	Z Lin (2002). Discovering EFL learner's perception of prior knowledge and its roles... J RES READ 25(2), 172-90\\
10	&	K HUIE (2003). Learning to write in the Primary Grades: Experiences of... TESOL JOURNAL 12(1), 25-31\\\hline
\end{tabular}
\end{center}
\caption{\label{recomm2} Usage-based recommendations for "Teaching reading to early language learners" query.}
\end{table*}

\section{Ongoing issues}

A number of noteworthy issues related to the large-scale aggregation and exploitation of usage data were encountered in the course of the reported work.  They are described in the remainder of this section.

\subsection{Linking server: representativeness}
There are drawbacks associated with the use of linking servers in a usage log aggregation framework relating to scope, scale and representativeness.  Indeed, although OpenURL is widely supported by scholarly information services, support is not universal, and especially new types of nodes in the scholarly communication environment such as Institutional Repositories and Dataset Repositories lag behind. Also, information services present value-added services to users, the use of which is only recorded at the level of the information service itself, not at the level of the linking server.  As a result, the linking server logs do not capture all events related to documents referenced in information services. Indeed, linking server logs may validly represent the actions of its user base, but they will inevitably miss certain aspects of usage. Future investigations need to focus on the definition of sampling statistics to determine the representativeness of linking server logs.

\subsection{Referent deduplication}

When linking server logs are recorded and aggregated, it is of vital importance that usage events pertaining to the same or a different {\it Referent} are recognized as such, i.e. the aggregated usage data must be de-duplicated at the level of the {\it Referents} to avoid over- and undercounting which occurs when two {\it Referents} are falsely confounded or distinguished respectively.

The issue of {\it Referent} de-duplication was approached by the introduction of a metadata-based de-duplication key that met the following criteria:

\begin{enumerate}
\item The metadata components used in the construction of the de-duplication key must be available for a large majority of the processed {\it Referents}. If not, many events would end up with empty fields in their keys and hence would lead to problematic de-duplication results.
\item A maximum number of identical {\it Referents} and a minimum number of dissimilar {\it Referent} should be joined.
\end{enumerate}

To identify de-duplication key candidates, we adopted an iterative procedure which selected those n-tuplets of metadata items which occurred in the highest number of {\it Referents}. From these candidates a final key was selected which offered the best pragmatic compromise of the availability of metadata components and de-duplication results. The key consisted of:\\
{\scriptsize \texttt{\{issn, start\_page, publication\_year, M(article\_title, 25)\}}}\\
where {\scriptsize \texttt{M(article\_title, 25)}} represents a fuzzy match (Levenshtein distance) on the first 25 characters of the article title.

It should be noted that the proposed architecture does not depend on the simple de-duplication approach described here, and that alternative, and superior, schemes for the de-duplication of {\it Referents} can easily be integrated. Given the impact that the de-duplication process has on the aforementioned services, it is imperative that this remains an important topic for future research.

\subsection{Agent deduplication}

\begin{figure}[ht!]
\begin{center}
\includegraphics[width=3.2in]{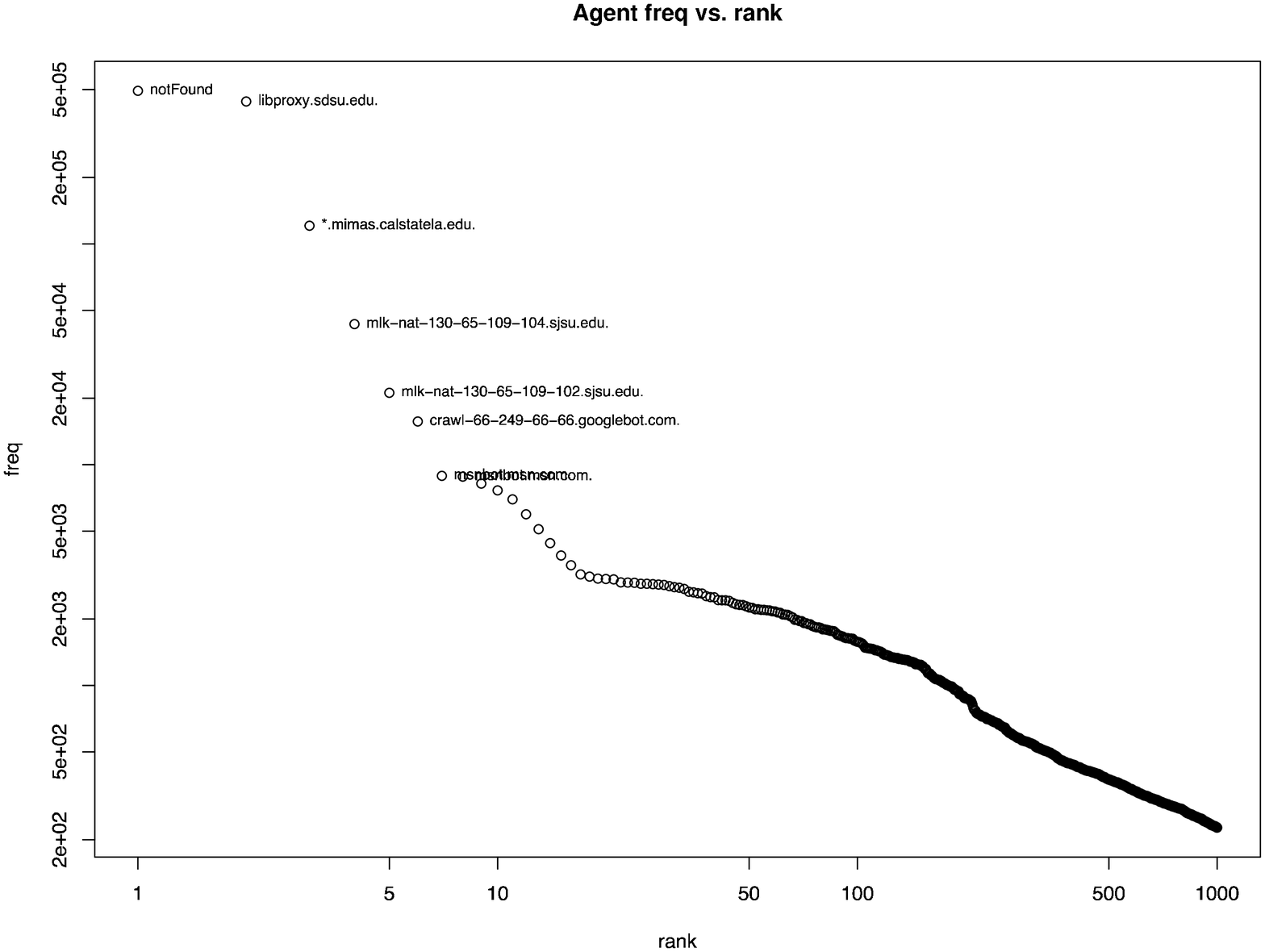}
\end{center}
\caption{\label{agentfreq} Distribution of IP addresses frequency in CalState logs. }
\end{figure}

The use of IP addresses to identify {\it Requesters} is prevalent, but leads to noisy usage data due to the use of proxies, {\it localhost} request, and robots/crawlers\footnote{Privacy concerns are discussed in the following section.}. Fig. \ref{agentfreq} shows how the distribution of the frequency of the request issued by particular IP addresses is distorted by the use of proxies and web robots and crawlers. The distribution follows a power-law only when the first 25 IP addresses are discarded. These IP addresses were indeed shown to correspond to {\it localhost} requests, proxies and robots/crawlers.

Three options were identified to mitigate this problem. First, the contributions of usage events originating from particular {\it Requesters} could be weighted inversely by their frequency of occurrence, i.e. the more frequent an IP address in the usage data, the lesser its contributions to the final usage statistics. This solution has the advantage that no predefined, manual filtering of usage data is required. Second, a manual filtering based on knowledge of the local linking server setup could be conducted. This is a highly effective approach but cumbersome, and not scalable because of the manual intervention. Third, a retooling of the linking server to use random, unique and anonymous session ID cookies rather than IP addresses could be adopted. Although this solution is prevalent in the Web-based retail environment, it still raises some controversy in the realm of scholarly information services because it relies on the use of client-side information which raises privacy concerns. 
\vspace{0.5cm}
\subsection{User privacy}

The collection and aggregation of usage data raises a multitude of legal and policy issues which became apparent in the reported development and evaluation. The foremost issue is that of user identification which has acquired a definite relevance in light of the recent demands placed on major search engines for reporting user actions. Although the proposed architecture allows extensive user and institutional data to be represented, this is not a requirement. Measures of user identity protection can be adopted both on the intra- and inter-institutional level, and accommodated by the proposed architecture. In particular, the use of anonymous, random user (or session) IDs to replace IP addresses has been explored.
In addition, modifications to the SFX linking server have been proposed to allow the use of such anonymous, session IDs. Future research will need to focus on the definition of approaches to further reduce the exposure of sensitive user- and institution related information.\\

\section{Conclusion}

We outlined an architecture for the large-scale recording, representation and aggregation of DL usage data. This architecture is  standards-based and relies on the already large base of installed linking servers and the wide-spread adoption of OpenURL and OAI-PMH based services. We discussed a recent evaluation of the architecture involving usage data recorded for the entire CalState system during the 2004 and 2005 period. It has been demonstrated that such logs provide an attractive starting point for services which support the end-user's scholarly activities, such as recommender systems, but likewise allow the scholarly community to monitor usage at a high level of detail.

The discussed evaluation of the proposed architecture is based on a particular configuration of usage data sources, a particular aggregation model and a particular set of service provisions. It is therefore limited in its generalizability.  Future research needs to be directed towards an investigation of the scalability of different aggregation architectures, models to protect user privacy, mechanisms to ensure data validity and detect fraud, metrics to determine data representativeness and a range of technical issues associated with {\it referent} identification. In addition, a range of potential applications of usage data can be further explored, i.e.~metrics of item impact, indicators of scholarly trends \cite{visual:chi1998} and end-user services beyond the recommender service discussed in this paper.

Such a wide range of technical, scientific and policy issues is associated with applications of usage data that they can not be addressed within the framework of a single research project. A community effort is required to fully explore this emerging domain. A community of scholars is slowly emerging but needs to be further consolidated. For that reason, the authors seek to organize a meeting of the different stakeholders and scholars in this area with the support of a prominent funding agency. The objective of such a meeting will be the establishment of a common research agenda around which a science of usage data can coalesce and develop.

\section*{Acknowledgements}

We thank Ex Libris who played an enabling role in our research and development efforts, as well as Marvin Pollard at California State University for his support and collaboration.


\end{document}